\documentclass{PoS}

\usepackage{amsmath}
\usepackage{microtype}

\title{Pion--nucleus Drell--Yan data as a novel constraint for nuclear PDFs}

\ShortTitle{Pion--nucleus DY data as a novel constraint for nPDFs}

\author{\speaker{Petja Paakkinen}
        \\ University of Jyvaskyla, Department of Physics, P.O. Box 35, FI-40014 University of Jyvaskyla, Finland \\
        E-mail: \email{petja.paakkinen@jyu.fi}}

\author{Kari J. Eskola\\
       University of Jyvaskyla, Department of Physics, P.O. Box 35, FI-40014 University of Jyvaskyla, Finland \\
       Helsinki Institute of Physics, P.O. Box 64, FI-00014 University of Helsinki, Finland \\
       E-mail: \email{kari.eskola@jyu.fi}}

\author{Hannu Paukkunen\\
       University of Jyvaskyla, Department of Physics, P.O. Box 35, FI-40014 University of Jyvaskyla, Finland \\
       Helsinki Institute of Physics, P.O. Box 64, FI-00014 University of Helsinki, Finland \\
       Instituto Galego de F\'{\i}sica de Altas Enerx\'{\i}as (IGFAE), Universidade de Santiago de Compostela, E-15782 Galicia, Spain \\
       E-mail: \email{hannu.paukkunen@jyu.fi}}

\abstract{We have studied the prospects of using the Drell--Yan dilepton process in pion--nucleus collisions as a novel input in the global analysis of nuclear parton distribution functions (nPDFs). In a NLO QCD framework, we find the measured nuclear cross-section ratios from the NA3, NA10 and E615 experiments to be largely insensitive to the pion parton distributions and also compatible with the EPS09 and nCTEQ15 nPDFs. These data sets can thus be, and in EPPS16 have been, included in global nPDF analyses without introducing significant new theoretical uncertainties or tension with the other data. In particular, we explore the constraining power of these data sets on the possible flavour asymmetry in the valence-quark nuclear modifications. Moreover, using the COMPASS kinematics we present predictions for pion charge-difference ratio, a new measurable which could help to further constrain this asymmetry.}

\FullConference{XXV International Workshop on Deep-Inelastic Scattering and Related Subjects\\
		3-7 April 2017\\
		University of Birmingham, UK}

\begin{document}

\vspace{-0.08cm}
\section{Introduction}
\vspace{-0.08cm}

An open and topical subject in the field of nuclear parton distribution functions (nPDFs) is the flavour dependence of quark nuclear modifications. In the past, due to lack of constraining data, most analyses assumed identical modifications for valence quarks (and separately for light sea quarks) at the parametrization scale. While no conflict with this assumption has been observed, the amount of allowed flavour asymmetry in the quark distributions is of particular interest e.g.\ when making predictions for observables such as the electroweak boson production at the LHC, which are sensitive to this asymmetry.

Here, we review the findings of our study~\cite{Paakkinen:2016wxk} based on the EPS09~\cite{Eskola:2009uj} and nCTEQ15~\cite{Kovarik:2015cma} nPDFs on the prospects of using pion--nucleus Drell--Yan (DY) dilepton data to constrain the flavour dependence.
We also show comparisons with the recent EPPS16~\cite{Eskola:2016oht} fit, where the considered DY data have been used as an input.
In addition, we propose a new observable, a pion charge-difference ratio, which promises to have a good sensitivity to the flavour separation of valence modifications.

\vspace{-0.08cm}
\section{Applicability of the available data sets}
\vspace{-0.08cm}

\begin{floatingfigure}
  \includegraphics[width=8.3cm]{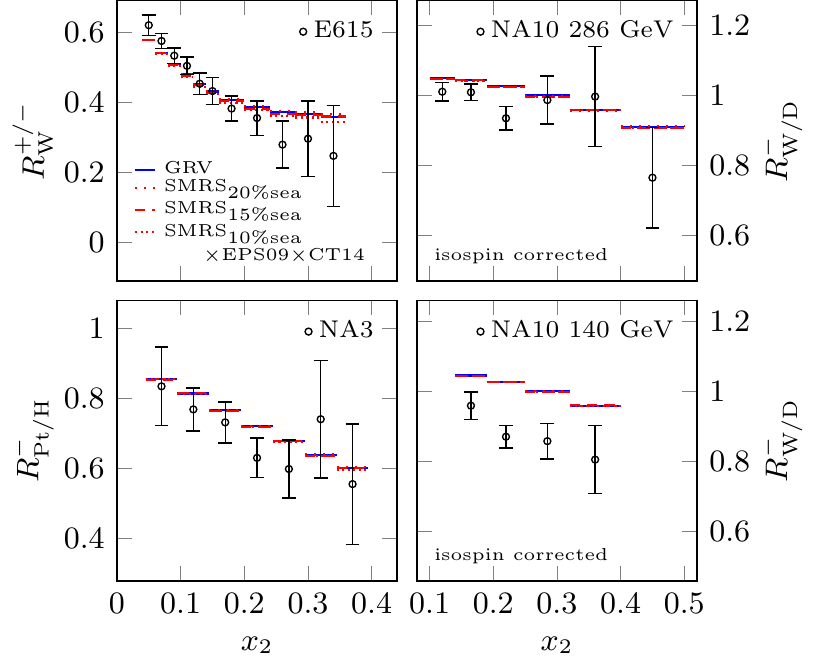}
  \vspace{-0.16cm}
  \caption{Comparison of full NLO calculations using the GRV and SMRS pion PDFs. No significant differences between the sets are observed. Figure from Ref.~\cite{Paakkinen:2016wxk}.}
  \label{fig:pion-pdf-comparison}
\end{floatingfigure}

We consider here the following nuclear cross-section ratios, differential in $x_2 \equiv \frac{M}{\sqrt{s}} \mathrm{e}^{-y}$, where $M$, $y$ are the invariant mass and rapidity of the lepton pair,
\begin{equation}
  \begin{split}
    R^{+/-}_A(x_2) &\equiv \frac{\mathrm{d}\sigma^{\pi^+ + A}_\text{DY} / \mathrm{d}x_2}{\mathrm{d}\sigma^{\pi^- + A}_\text{DY} / \mathrm{d}x_2}, \\
    R^{-}_{A_1/A_2}(x_2) &\equiv \frac{\frac{1}{A_1} \mathrm{d}\sigma^{\pi^- + A_1}_\text{DY} / \mathrm{d}x_2}{\frac{1}{A_2} \mathrm{d}\sigma^{\pi^- + A_2}_\text{DY} / \mathrm{d}x_2},
  \end{split}
\end{equation}
as provided by the NA3~\cite{Badier:1981ci}, NA10~\cite{Bordalo:1987cs} and E615~\cite{Heinrich:1989cp} experiments.
Assuming isospin (IS) and charge-conjugation (CC) symmetry between $\pi^+$ and $\pi^-$, we see that the quark distributions in charged pions are related with
$u_{\pi^{+}} \overset{\text{IS}}{=} d_{\pi^{-}} \overset{\text{CC}}{=} \bar{d}_{\pi^{+}} \overset{\text{IS}}{=} \bar{u}_{\pi^{-}}$
and
$d_{\pi^{+}} \overset{\text{IS}}{=} u_{\pi^{-}} \overset{\text{CC}}{=} \bar{u}_{\pi^{+}} \overset{\text{IS}}{=} \bar{d}_{\pi^{-}}$.
Now, in the kinematical limit where the pion sea quarks can be neglected, the leading order (LO) approximation for a narrow-enough invariant mass bin gives
\begin{equation}
  R^{+/-}_A(x_2) \approx \frac{4\bar{u}_A(x_2) + d_A(x_2)}{4u_A(x_2) + \bar{d}_A(x_2)}, \qquad
  R^{-}_{A_1/A_2}(x_2) \approx \frac{4u_{A_1}(x_2) + \bar{d}_{A_1}(x_2)}{4u_{A_2}(x_2) + \bar{d}_{A_2}(x_2)},
  \label{eq:approximation}
\end{equation}
i.e.\ the dependence on pion PDFs essentially cancels in the above ratios~\cite{Dutta:2010pg}. We have verified that this cancellation works well also at the next-to-leading order (NLO) level. This can be seen from Figure~\ref{fig:pion-pdf-comparison}, where the results from the GRV~\cite{Gluck:1991ey} and SMRS~\cite{Sutton:1991ay} pion PDFs are compared. This is important, as it indicates that these data can be included in a global nPDF fit without imposing significant new theoretical uncertainties from the pion structure.

\begin{floatingfigure}
  \includegraphics[width=8.3cm]{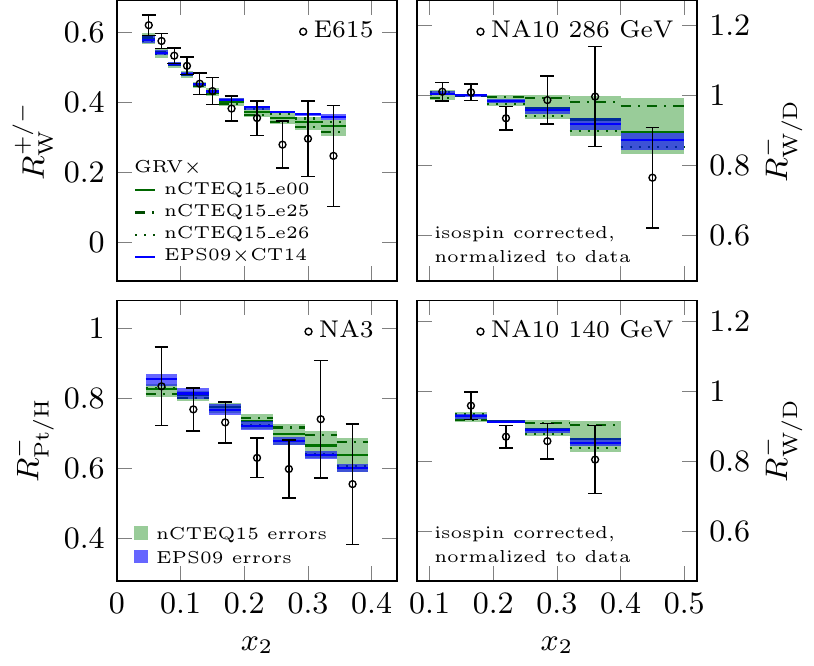}
  \vspace{-0.16cm}
  \caption{Results with the EPS09 and nCTEQ15 nPDFs~\cite{Paakkinen:2016wxk}. The EPS09 results are obtained using the CT14~\cite{Dulat:2015mca} proton baseline PDFs. We show here also the predictions with the nCTEQ15 error sets 25 and 26.}
  \label{fig:nuclear-pdf-comparison}
\end{floatingfigure}

For data-to-theory comparison, one has to take into account the isospin correction and the systematic overall normalization uncertainty in the NA10 data sets. The technical details on this matter can be found in Ref.~\cite{Paakkinen:2016wxk}, we simply note here that after correcting the NLO predictions with
\begin{equation}
  \begin{split}
    &(R^{-}_\mathrm{W/D})^\text{NLO}_\text{isospin corrected} \\&= (R^{-}_\text{isocalar-$\mathrm{W/W}$})^\text{LO}_\text{no nPDFs} \times (R^{-}_\mathrm{W/D})^\text{NLO}
  \end{split}
\end{equation}
and accounting for the data normalization uncertainty (``normalizing to data''), both EPS09 and nCTEQ15 are in a good agreement with the measurements, as is evident from Figure~\ref{fig:nuclear-pdf-comparison}, indicating that here is no tension between these measurements and other data used in the two analyses.\footnote{This comes with a side note that although we find the higher beam energy predictions for NA10 to be within the given $6\%$ normalization uncertainty interval, for the lower energy we need $\sim 12\%$ correction.}

\vspace{-0.08cm}
\section{Comparison of nPDF results}
\vspace{-0.08cm}

While the data are well described by both EPS09 and nCTEQ15, these nPDFs have large differences in their uncertainty estimates. To understand where this comes from, we have plotted in Figure~\ref{fig:nuclear-pdf-comparison} also the predictions with the nCTEQ15 error sets 25 ($R^{A}_{u_\text{V}} \ll R^{A}_{d_\text{V}}$) and 26 ($R^{A}_{u_\text{V}} \sim R^{A}_{d_\text{V}}$).
Here $ R^{A}_{i}(x, Q^2) = {f^{p/A}_{i}(x, Q^2)} / {f^{p}_{i}(x, Q^2)} $ is the nuclear modification of the distribution of a parton flavour $i$ in a bound proton in nucleus $A$ compared to that of a free proton.
The clear separation in the predictions with these two sets shows that the studied observables are sensitive to mutual differences in valence quark nuclear modifications. This is best understood in the context of the $R^{-}_\text{W/D}$ ratio measured by NA10. For large $x_2$ only valence quarks in nuclei contribute and in the LO approximation we have
\begin{equation}
  R^{-}_{A/\text{D}} \overset{x_2 \rightarrow 1}{\approx} \frac{u^\text{V}_{p/A} + d^\text{V}_{p/A}}{u^\text{V}_{p} + d^\text{V}_{p}} + \left(\frac{2Z}{A} - 1\right)\frac{u^\text{V}_{p/A} - d^\text{V}_{p/A}}{u^\text{V}_{p} + d^\text{V}_{p}}.
  \label{eq:rminusisospin}
\end{equation}
Here, the first term in the sum is the nuclear modification of an average valence quark in an isoscalar nucleus. The sensitivity to the valence asymmetry comes from the second term and is limited by the amount of neutron excess (non-isoscalarity) in the nucleus.

We find the nCTEQ15 error bands to be large since in their analysis the flavour dependence was allowed, but not well constrained. Conversely, the EPS09 error sets underestimate the true uncertainty because the flavour dependence of valence quark nuclear modifications was not allowed.
We also observe that the predictions with the nCTEQ15 error set 25 do not reproduce the slope of the NA10 data particularly well. This indicates such a large asymmetry to be unlikely.

\begin{floatingfigure}
  \includegraphics[width=8.3cm]{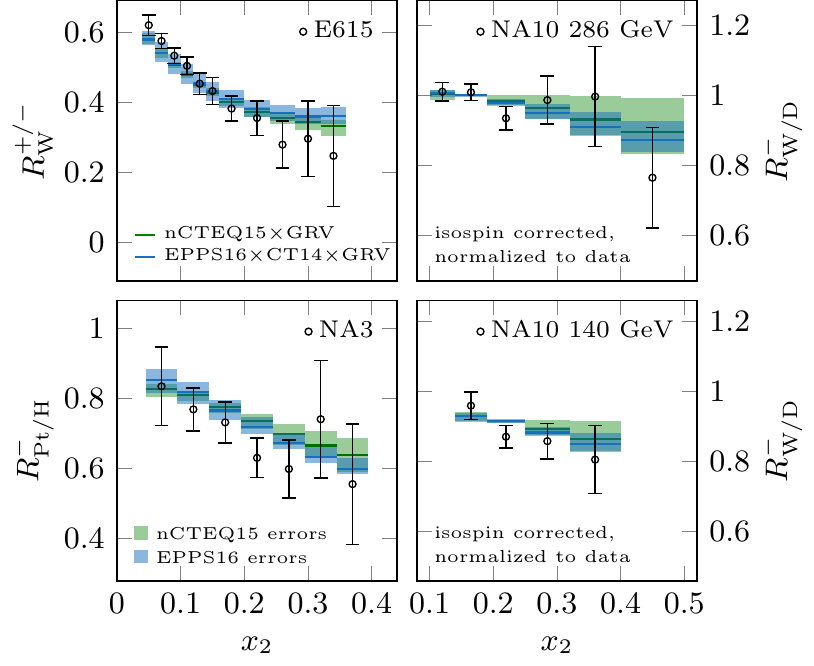}
  \vspace{-0.16cm}
  \caption{The EPPS16 results using the CT14 baseline compared with nCTEQ15.}
  \label{fig:nuclear-pdf-comparison-2}
\end{floatingfigure}

Motivated by the above results, these data were used in the new EPPS16 analysis, where both valence flavours were allowed to vary independently. The results, as shown in Figure~\ref{fig:nuclear-pdf-comparison-2}, are similar to EPS09, but with larger errors,  as is expected for having more freedom in the fit. Compared to nCTEQ15 there is a reduction in error estimates. This, however, is not due to the pion--nucleus Drell--Yan data, as at the moment more stringent constraints come from neutrino-induced deep inelastic scattering (DIS) together with the proper treatment of isoscalar corrections in charged-lepton DIS data, as explained in Ref.~\cite{Eskola:2016oht}. Thus, while the DY observables considered here could in principle constrain the valence asymmetry, the available data is not precise enough for this.

\vspace{-0.08cm}
\section{New observable}
\vspace{-0.08cm}

The approximation in Equation~\eqref{eq:approximation} required us to be in a kinematical region where pion sea quarks do not give a significant contribution to the cross-section (i.e. at large $x_1 \equiv \frac{M}{\sqrt{s}} \mathrm{e}^{y}$). This restriction can be avoided by considering the ratio of the \emph{difference} of the negative and positive charged pion cross-sections
\begin{equation}
  R^\Delta_{A_1/A_2}(x_2) \equiv \frac{\frac{1}{A_1} (\mathrm{d}\sigma^{\pi^- + A_1}_\text{DY} / \mathrm{d}x_2 - \mathrm{d}\sigma^{\pi^+ + A_1}_\text{DY} / \mathrm{d}x_2)}{\frac{1}{A_2} (\mathrm{d}\sigma^{\pi^- + A_2}_\text{DY} / \mathrm{d}x_2 - \mathrm{d}\sigma^{\pi^+ + A_2}_\text{DY} / \mathrm{d}x_2)}.
\end{equation}
In LO all sea quark contributions cancel, and hence this ratio depends only on nuclear valence distributions
\begin{equation}
  R^\Delta_{A_1/A_2}(x_2) \approx \frac{4u^\mathrm{V}_{A_1}(x_2) - d^\mathrm{V}_{A_1}(x_2)}{4u^\mathrm{V}_{A_2}(x_2) - d^\mathrm{V}_{A_2}(x_2)}.
\end{equation}
When $A_2 = \text{D}$, we can write this as
\begin{equation}
  R^\Delta_{A_1/\text{D}} \approx \frac{u^\mathrm{V}_{p/A} + d^\mathrm{V}_{p/A}}{u^\mathrm{V}_p + d^\mathrm{V}_p} + \frac{5}{3} \left(\frac{2Z}{A} - 1\right) \frac{u^\mathrm{V}_{p/A} - d^\mathrm{V}_{p/A}}{u^\mathrm{V}_p + d^\mathrm{V}_p},
\end{equation}
where we notice a factor $5/3$ increase in the non-isoscalar part compared to Equation~\eqref{eq:rminusisospin}. This promises an enhanced sensitivity to the valence asymmetry. In Figure~\ref{fig:newobservable} we plot predictions for the suggested pion charge-difference ratio with beam energy and acceptances available at the COMPASS experiment~\cite{Gautheron:2010wva,Abbon:2014aex}. Indeed, in contrast to EPS09, for which $R^{A}_{u_\text{V}} \approx R^{A}_{d_\text{V}}$ by construction, we find large error bands for both EPPS16 and nCTEQ15; a measurement could help to reduce these.

\begin{floatingfigure}
  \includegraphics[width=11.5cm]{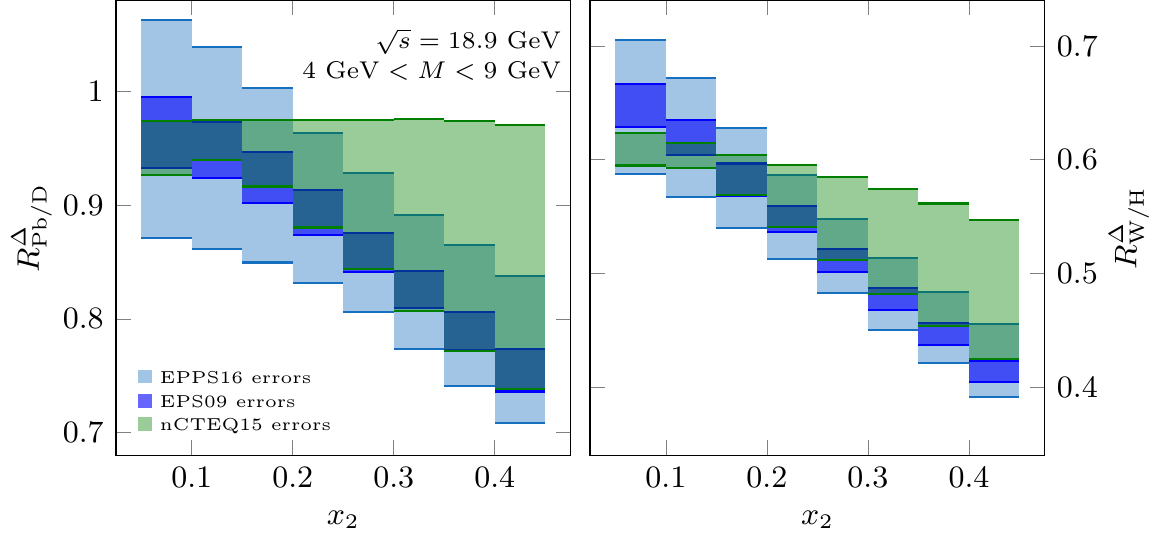}
  \vspace{-0.16cm}
  \caption{Predictions for the pion charge-difference ratio using the COMPASS kinematics.}
  \label{fig:newobservable}
\end{floatingfigure}

\vspace{-0.08cm}
\section{Summary}
\vspace{-0.16cm}

We have considered here the prospects of using the Drell--Yan dilepton process in pion--nucleus collisions as a novel input in the global analysis of nPDFs and the possible impact on flavour asymmetry of valence quarks.
We have found the data from the NA3, NA10 and E615 experiments to be compatible with modern nPDFs. These data can thus be used in global analyses without causing significant tension with other data. This has been recently realized in the EPPS16 analysis.
The cross-section ratios are largely independent of the pion PDFs and hence the inclusion of these data in global nPDFs fits does not impose significant new biases.
While we find the available data to be consistent with flavour-symmetric valence modifications, the statistical precision is not high enough to give meaningful constraints for the asymmetry.
To this end, we propose a new observable to be measured.

\vspace{-0.32cm}
\acknowledgments
\vspace{-0.24cm}

The authors have received funding from Academy of Finland, Project 297058 of K.J.E; the European Research Council grant HotLHC ERC-2011-StG-279579; Ministerio de Ciencia e Innovaci\'{o}n of Spain and FEDER, project FPA2014-58293-C2-1-P; Xunta de Galicia (Conselleria de Educacion) - H.P.\ is part of the Strategic Unit AGRUP2015/11. P.P.\ acknowledges the financial support from the Magnus Ehrnrooth Foundation.

\end{document}